\documentclass{emulateapj}
\usepackage{graphicx,amsmath}
\usepackage{amsfonts}

\usepackage{natbib}
\slugcomment{}
\shorttitle{mm-wave strong lensing}
\shortauthors{Hezaveh, Holder}
\begin{document}
\title{Effects of Strong Gravitational Lensing\\
 			on millimeter-wave Galaxy Number Counts} 			
\author{Yashar D. Hezaveh, Gilbert P. Holder}
\affil{Department of Physics, McGill University,
    Montreal, QC, Canada}
    
\begin{abstract}
We study the effects of strong lensing on the observed number counts of mm sources using a ray tracing simulation and two number count models of 
unlensed sources. We employ a quantitative treatment of maximum attainable 
magnification factor depending on the physical size of the sources, also
accounting for effects of lens halo ellipticity. 
We calculate predicted number counts and redshift distributions 
of mm galaxies including the effects of strong lensing and compare with
the recent source count measurements of the South Pole Telescope (SPT).
The predictions have large uncertaities, especially the details of the
mass distribution in lens galaxies and the finite extent of sources, but
the SPT observations are in good agreement with predictions. 
The sources detected by SPT are predicted to largely consist of strongly 
lensed galaxies at $z>2$. The typical magnifications of these sources 
strongly depends on both the assumed unlensed source counts and the
flux of the observed sources. 
\end{abstract}

%\keywords{gravitational lensing ---
%galaxies: luminosity function, mass function---
%galaxies: abundances---
%methods: numerical
%}



\section{Introduction}

Many galaxies at redshifts $z \sim 2-5$ have been found to be
undergoing large amounts of star formation, leading to a population
of distant galaxies with large amounts of warm dust that can be
observed at infrared, submm, and mm wavelengths \citep{Blain:02}. 
Star-forming rates are found to be often in excess of
~1000 $M_\odot$/yr 
\citep{Michalowski:10}, contributing a significant fraction of the
total cosmic star formation at these redshifts.

Surveys at submm wavelengths \citep{Coppin:06} covering smaller areas at high
sensitivity have established the existence of a population of
dusty star-forming galaxies at high redshift \citep{Blain:02}, 
showing that the number of sources as a function of flux (the luminosity
function) is steeply declining at high fluxes. As a result, gravitational
lensing is expected to significantly modify the observed number counts
\citep{Blain:98}, an effect known as ``magnification'' or
``amplification'' bias \citep{Turner:84}. 
Recent mm-wave surveys like the South Pole Telescope (SPT; Vieira et al
2010)\nocite{Vieira:10}  are now covering enough area to accumulate
statistically significant numbers of highly luminous distant galaxies,
providing an opportunity to compile large samples of strong gravitational
lenses. 
Recent theoretical work \citep{Negrello:07, Fedeli:09, Lima:09, Lima:10, Jain:10} 
has demonstrated that gravitational lensing
is likely an important contributor to the galaxy counts observed by large
scale mm-wave surveys. 
In addition, evidence is now emerging from
Herschel observations \citep{Frayer:10} that a large fraction of
the brightest high-redshift dusty galaxies are indeed strongly lensed. 

Much remains unknown about massively star-forming galaxies at high
redshift. The redshift distribution as a function of flux is only
roughly understood, and different models have very different
input physics. For example, the Durham semi-analytic model of galaxy
formation requires a top-heavy IMF to explain this population
\citep{Lacey:10}. Strong lensing of these sources allows a magnified view,
making multi-wavelength follow-up easier, as the sources are brighter.

In this work we calculate the expected number of strongly
lensed galaxies in flux-limited mm-wave surveys, paying particular 
attention to the expected redshift distribution of the sources and
lenses and the effect of finite source effects. 

\section{Overview of Calculation of Number of Lensed Sources}

Determining the expected number of galaxies discovered in mm-wave 
surveys is complicated by at least four major uncertainties:
\begin{itemize} 
\item the statistics of the source
population (uncertain number counts, uncertain redshift distribution)
\item the properties of the source population (uncertain spectral energy
distributions, uncertain angular sizes)
\item the statistics of the lens
population (number counts as a function of mass and redshift)
\item the properties of the lens galaxies (internal mass profiles
and ellipticities).
\end{itemize}

To investigate uncertainties in the source population, 
we use two independent unlensed source count predictions for SPT measurements
at 220 GHz (1.4 mm).
In particular, the redshift distribution is expected to play
a key role in determining lensing efficiencies, so the two different
models are intended to provide an estimate of the redshift importance. 
Our first model, henceforth called the \textit{Durham model}, based
on the models developed in \citet{Baugh:05} 
(Lacey et al. private communication), is the result of semi-analytic modeling of galaxy formation.
The second model considered, referred to as the
\textit{UBC model} \citep{Marsden:10},
 is obtained through backward evolution models of the local Universe.

To model the lens population we follow \citet{Perrotta:02}, using
a Press-Schechter \citep{Press:74} approach to the lens halo distribution
as a function of mass and redshift. 
We use the Sheth \& Tormen (1999) redshift-dependent mass function to model
the number of halos of a given mass and redshift for our lens population. 

For the internal mass distribution
we assume that the region where the majority of strong lensing occurs can
be modeled as an elliptical mass profile with a 3D density profile
that falls as $1/r^2$. This is an excellent approximation for galaxies
\citep{Koopmans:09}, while it is likely not a sufficiently complex model
to capture the lensing properties of massive galaxy clusters \citep{Richard:10}.

We use ray-tracing simulations to explore the impact of 
lens ellipticity and finite source sizes, assuming constant values of
ellipticity and source sizes and exploring the impact of different 
assumed values.

In all that follows, we assume as our fiducial cosmology 
a spatially flat universe with $\Omega_m=0.222$, $H_\circ=71.0 \,{\rm km\,s^{-1}\, Mpc^{-1}}$,
$\Omega_b=0.0449$, $n_s=0.963$ and $\sigma_8=0.801$.

\section{Unlensed mm-Wave Number Count Predictions}

The assumed unlensed source count
models (the Durham and UBC models) have not been
calibrated at mm wavelengths; small differences in parameters such as
dust emissivity that are not large effects at submm wavelengths could lead
to large misestimates at mm wavelengths. 

A simple check is to verify that the models produce a reasonable amount
of noise power in mm-wave maps; this has been measured in SPT data
in \citet{Hall:09}.
We computed the angular noise power, defined for randomly distributed 
point sources as
\begin{equation}
c_l = \int_{0}^{S_{cut}} \: S^3 \: \frac{dN}{dS} \: d(ln S)
\label{powerEQ}
\end{equation}
with a flux cutoff of 17 mJy \citep{Hall:09} for SPT at 220 GHz (1.4 mm). 
The majority
of the noise power comes from sources well below the SPT sensitivity 
limit for detecting individual sources. 

\begin{figure}[h]
\centering
\makebox[0cm]{\includegraphics[trim=80mm 4mm 80mm 4mm, clip, scale=0.48]{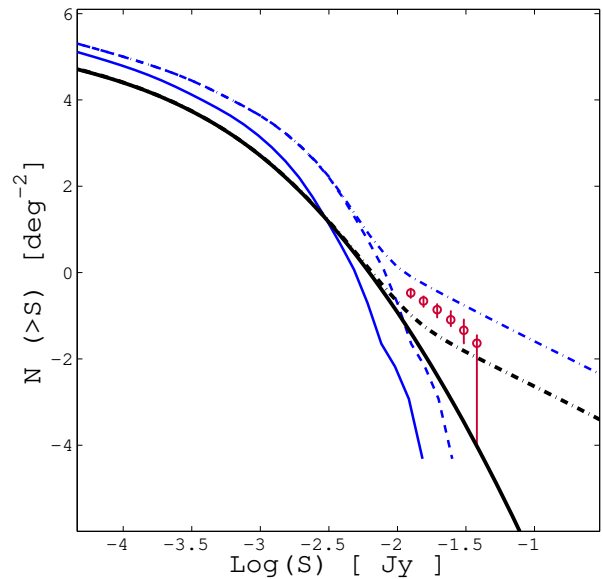}}
\caption{Unlensed number count predictions for SPT sources at 220 GHz. 
The thin dash-dotted blue line shows the original Durham counts while the thin dashed 
blue curve shows the same counts with $z>0.2$. The thin solid blue curve is 
the Durham model corrected to match the measured noise power \citep{Hall:09}. 
The thick black dash-dotted line shows the total UBC counts and the thick solid black line is UBC with $z>0.2$ for comparison purposes. The data points are 220 GHz 
(1.4 mm) SPT
dusty galaxies after removing known IRAS sources \citep{Vieira:10}.}
\label{f1}
\end{figure}

The noise power from the UBC model is in excellent agreement with the
measurements, and did not require any corrections. 
The noise power derived from the Durham model, around 60 Jy$^2$/sr,
is more than a factor of 3 too high compared to \citet{Hall:09}. 
We scale the flux of each object by a constant amount to match the 
measured noise power.  In Figure \ref{f1} it can be seen that the
Durham and UBC models are forced by this constraint on the total power
to have comparable number counts at fluxes of a few mJy.

After applying a constant flux correction to match the observed noise
power, the Durham model still had a problem with the properties of 
low-redshift dusty galaxies, in that the predicted number of low-redshift
galaxies in the SPT sample was too high. In particular, the brightest of
the low-z population
in the Durham model (observed by IRAS) are predicted to be bright enough
in the SPT maps to be found as sources. This is not the case, as evidenced
by the small fraction of SPT-discovered galaxies that were also observed to
be IRAS sources \citep{Vieira:10}\footnote{http://pole.uchicago.edu/public/data/vieira09/index.html}. 
As gravitational lensing is more efficient for high redshift
sources, we elected to simply further suppress the flux of low-z  galaxies
($z<0.2$) to make the low-z galaxy counts agree with the number of SPT
sources found to coincide with IRAS sources. While not rigorously 
justifiable, the main point of using the Durham model was to get a
plausible redshift distribution of sources at high-redshift. In order to avoid misinterpretations caused by the low-z population we only apply our lensing model to Durham counts with $z>0.2$ eliminating the low redshift counts and compare the results to SPT's IRAS removed counts.

\section{Gravitational Lensing Theory}

For the details of gravitational lensing theory we refer the reader to a review by \citet{Bartelmann:01}. Here we only briefly state a few lensing quantities that are used in this work. The lens equation that we solve numerically is written as
\begin{equation}
\vec{\beta}=\vec{\theta}-\frac{D_{ds}}{D_s}\hat{\vec{\alpha}}(D_d \vec{\theta})=\vec{\theta}-\vec{\alpha}(\vec{\theta})
\label{lensEQ}
\end{equation}
where $D_d$, $D_s$, and $D_{ds}$ are the angular diameter distances of deflector (d) and source (s) and $\vec{\beta}$ is the observed position of a point source at $\vec{\theta}$ deflected by an angle $\vec{\alpha}$. The total 
deflection from an ensemble of point masses at a single lens plane is given by
\begin{equation}
\hat{\vec{\alpha}}(\vec{\theta})=\frac{1}{\pi}\int d^2\theta'\kappa(\vec{\theta}')\frac{\vec{\theta}-\vec{\theta'}}{|\vec{\theta}-\vec{\theta'}|^2}
\label{lens equation}
\end{equation}

The dimensionless surface mass density $\kappa$ is
\begin{equation}\begin{split}
\kappa(\vec{\theta})=\frac{\Sigma(D_d\vec{\theta})}{\Sigma_{cr}}  \quad \mbox{where} \quad \Sigma_{cr}=\frac{c^2}{4\pi G}\frac{D_s}{D_d D_{ds}}
\end{split}\end{equation}
is the critical surface mass density and $\Sigma(\vec{\xi})$ is the 2D projected mass density of the lens.

\vspace{5 mm}

\section{Lens Modeling and Calculation Details}

\subsection{Ray-tracing}
The lens equation (Equation \ref{lensEQ}) is an implicit equation, 
in that the image position $\vec{\theta}$ is needed to evaluate
the deflection $\vec{\alpha}(\vec{\theta})$. For a given source 
position $\vec{\beta}$ it is difficult to solve this implicit equation to obtain $\vec{\theta}$. On the other hand, if $\vec{\theta}$ is known 
$\vec{\beta}$ can be easily evaluated. This constitutes the basis of ray-tracing
simulations of gravitational lensing: start with an array of image positions
and determine the source positions from which they emerged. 

Simple halo profiles, like the singular isothermal sphere (SIS), 
can often produce analytical cross-sections. 
More complex mass profiles often have no simple analytical solution 
and the cross-section has to be numerically computed using a ray 
tracing simulation. 
In addition to the ability to compute lensing quantities for any arbitrary mass configuration, ray-tracing simulations have the advantage that they 
can easily include finite source effects which are not properly 
modeled in analytical solutions.

Our simulation makes surface density maps as a matrix of $400\times 400$ 
elements and solves the lens equation using the method described in 
\citet{Keeton:01}. The image plane is divided to $180\times 180$ 
squares each of which is divided into two triangles and the corresponding
positions of each vertex are found in the source plane. 
The magnification is computed by defining a grid in the 
source plane and calculating the image-plane area of triangles that have been 
mapped into each source position.

\subsection{Halo Mass Profile and Ellipticity}

We assume that the lens profiles in the region of interest are 
well-approximated by singular isothermal profiles, where the
three dimensional density profile for a spherical profile is
\begin{equation}
\rho(r)=\frac{\sigma_{\nu}^2}{2\pi G r^2}
\end{equation}
where $\sigma_{\nu}$ is the line of sight velocity dispersion of the stars in the galactic disk or the galaxies in a galaxy cluster. 
It is well known that this is not a good approximation to halos produced
in dark matter simulations \citep{NFW:97}, but the region that dominates
the strong lensing properties is typically dominated by baryonic processes
and is empirically found to be close to isothermal \citep{Koopmans:09}.

The dependence of the velocity dispersion on the redshift and mass of the halo is given by \cite{Bryan:98} as
\begin{equation}
\sigma_{\nu}= f_{\sigma} M^{1/3}\left[\frac{H^2(z)\Delta(z)G^2}{16}\right]^{1/6}
\end{equation}
where
\begin{equation}
\Delta(z)=18\pi^2+82(\Omega_m-1)-39(\Omega_m-1)^2
\end{equation}
and $f_{\sigma}$ is a scaling parameter used to match the normalization from simulations. While the density profile may indeed scale roughly as
$1/r^2$ in the central regions of interest, 
the relationship between the normalization (i.e., the velocity dispersion) 
and total halo mass is not empirically calibrated and is a potential large 
source of systematic uncertainty in this analysis. 

Integrating $\rho(r)$ along the line-of-sight produces the projected surface mass density
\begin{equation}
\Sigma(\xi)= \frac{\sigma_{\nu}^2}{2 G \xi}
\end{equation}
The corresponding dimensionless surface mass density is
\begin{equation}\label{e:barwq}
\kappa(\theta)=\frac{\theta_E}{2\theta}  
\end{equation}
where we have defined the  Einstein deflection angle as
\begin{equation}
\theta_E=4\pi\left(\frac{\sigma_{\nu}}{c}\right)^2\frac{D_{ds}}{D_s}
\end{equation}

The magnification for point sources lensed by this mass profile is
analytic, with 
$ \mu(\vec{\theta})=|\vec{\theta}|/(|\vec{\theta}|-\theta_E) $.
We calculate magnification for extended sources as the 
ratio of the area of 
combined images to the source area. This allows easy exploration of
finite source effects and non-trivial lens distributions.

The integral lensing cross-section $\sigma(\mu>\mu_{min})$ is the area on the source plane inside which the magnification of a source is equal or 
larger than $\mu_{min}$. Throughout this work $\sigma(\mu>\mu_{min})$ and $\sigma(\mu)$ are used interchangeably.
The cross-section for an SIS halo for $\mu_{min}>2$ is given analytically by 
Perrotta et al. (2002) and at large magnifications our numerical results agree with the analytic
form to better than 2.5$\%$.

The morphology of the galaxies from observations \citep{Evans:09} and the shape of dark matter halos from N-body simulations \citep{Ludlow:10} both indicate that a considerable amount of ellipticity is present in the lensing halos. Many studies previously (e.g., \citet{Meneghetti:05}) have introduced ellipticity in the projected two-dimensional lensing potential $\phi$. For analytical studies this has the advantage that the second derivatives of the potential directly give the lensing quantities and lead to simple analytical expressions. However for high values of ellipticity this implies dumbbell shape density profiles which are unrealistic. 
Simulations and observations both indicate ellipticity in the distribution of 
mass, rather than in the potential resulting from it. In our ray-tracing 
approach it is simple to introduce ellipticity in the actual surface mass density. At each point defined by x and y on the two-dimensional plane with radius $r=\sqrt{x^2+y^2}$ we compute the SIS or NFW density using radius $r_e$ defined as
\begin{equation}
r_e=\sqrt{\frac{x^2}{1-e_{\kappa}}+y^2(1-e_{\kappa})}
\end{equation}
The axis ratios are now given by $b/a=1-e_{\kappa}$.

\begin{figure}[h]
\centering
\makebox[0cm]{\includegraphics[trim=10mm 1mm 1mm 1mm, clip, scale=0.45]{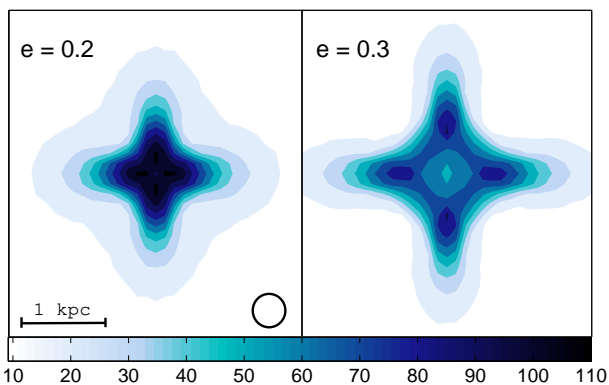}}
\caption{Magnification maps of two halos with identical total mass ($M=10^{13}M_{\odot}$) and 
different ellipticities ($e_{\kappa}=0.2$ for the left panel and $e_{\kappa}=0.3$ for the right panel). 
The extended circular source (drawn as a black circle) is identical in both plots making all magnification differences due to ellipticity. 
The left panel contains large magnifications in excess of $\mu=100$ which are damped in the right panel. But intermediate magnifications are extended to larger regions in the right panel.}
\label{f2}
\end{figure}

The impact of ellipticity can be clearly seen in Figure \ref{f2}.
Ellipticity increases the area of the source plane where strong lensing
can occur. However, in the same figure we can see the effects of
finite sources. The regions of high magnification become narrower
as ellipticity increases, so larger sources will tend to have smaller
peak magnifications.

\begin{figure}[h]
\centering
\makebox[0cm]{\includegraphics[trim=80mm 1mm 80mm 1mm, clip, scale=0.45]{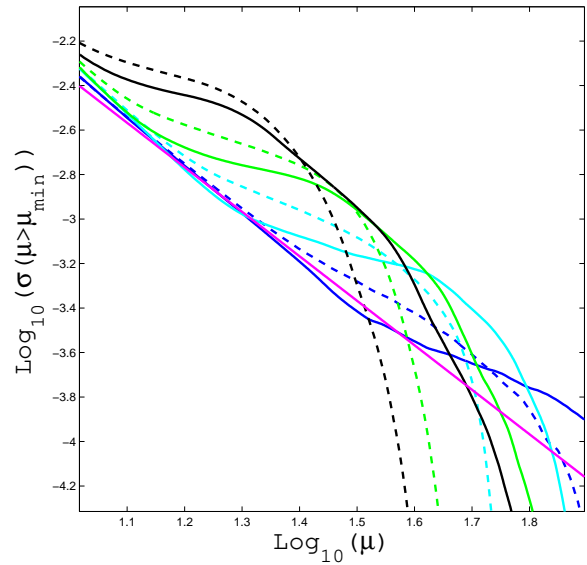}}
\caption{Above curves demonstrate the correlation of finite source effects with halo ellipticity. As halos become more elliptical they become more sensitive to finite source effects and dampen faster at high magnification. The colors correspond to different lens halo ellipticities: blue, cyan, green, and black correspond to $e_{\kappa}=0.1$, $e_{\kappa} = 0.2$, $e_{\kappa} = 0.3$, and $e_{\kappa} = 0.4$ respectively. Solid lines are the cross-sections for a source with a radius of 0.023$R_{Ein}$ while dashed lines correspond to a 0.041$R_{Ein}$ radius source.}
\label{f3}
\end{figure}

The interplay between ellipticity and source size is shown in Figure
\ref{f3}. Ellipticity generally leads to higher magnifications,
but finite source sizes become more important for higher ellipticities.
This discussion has assumed a fixed Einstein radius. The relevant factor
is the ratio of the source size to the Einstein radius; a given
source size that may be a problem for a galaxy-scale lens would 
behave like a point source for the purposes of lensing by a galaxy
cluster.

At intermediate magnifications, finite source effects tend to
increase the lensing cross-section. This occurs because sources are now
occupying a larger fraction of the source plane and are more likely to 
have a part of the emitting region be in the strong lensing regime.

%\begin{equation}
%\sigma(\mu>\mu_{min})=\frac{4\pi \hat{\alpha}^2 D_{ds}^2}{\mu_{min}^2}
%\label{cross-section}
%\end{equation}

\subsection{Probability function}

We use the halo mass function of \citet{Sheth:99} to determine the number
of halos of a given mass at each redshift, 
sum the source plane cross-sections of all lenses between the 
observer and source, and divide it by the total area of the sphere 
centered at the observer with a radius at the source redshift. 
This is the probability that a source at that redshift is magnified by at
least $\mu$ due to all intervening halos.

The sum of all cross-sections in a flat Universe can be written as
\begin{eqnarray}
 &\sigma_{tot}&(\mu,z_s,R_s,e_{\kappa})=4\pi\left(\frac{c}{H_o}\right) \int_0^{z_s} dz_d  \\
 &&\int dM\frac{\sigma(\mu,z_d,z_s,M,R_s,e_{\kappa}) n_c(z_d,M) (1+z_d)^2  D^2(z_d)}{\sqrt{\Omega_{oM}(1+z_d)^3+\Omega_{o\Lambda}}} \nonumber
\end{eqnarray}
 where $n_c(z_d,M)$ is the number density of lenses of mass $M$ at redshift $z_d$ and $D(z_d)$ is the angular diameter distance at redshift $z_d$. 
 
 Using the mass and redshift of our fiducial isothermal
halo profile, to find the cross-section for each lens as a function of mass M and redshift $z_d$ we can use the following scaling relation
\begin{eqnarray}
\sigma(\mu,z_d,z_s,M, R_s, e_\kappa)=&& \\
\left[\frac{M}{M_{0}}\right]^{4/3}&
\left[\frac{D_{ds}/D_s}{D_{ds,0}/D_{s,0}}\right]^2 &\chi^2 (z_d) \;  \sigma_{0}(\mu,\frac{R_s}{R'_{E}},\epsilon) \nonumber
\label{scaling}
\end{eqnarray}
where the subscript 0 refers to the quantities for a normalization halo from which the cross-section for other halos are achieved by the above scaling relation. $R'_E$ is the Einstein radius of the scaled halo and $\sigma_0$ is a look-up table of the cross-sections as a function of halo ellipticity and source-size to Einstein radius ratio. $\chi(z_d)$ is the scaling factor due to the dependence of the halo velocity dispersion, $\sigma_{\nu}$ on redshift and is given by
\begin{equation}
\chi (z_d)=\left(\frac{H^2(z_d)\; \Delta(z_d)}{H^2(z_{d,0}) \; \Delta(z_{d,0})}\right)^{1/3}
\end{equation}

The probability that a source at $z_s$ is magnified by a factor greater than $\mu_{min}$ is then
\begin{equation}
P(\mu>\mu_{min})=\frac{\sigma(\mu>\mu_{min})}{4\pi D_s^2}
\end{equation}

\begin{figure}[h]
\centering
\makebox[0cm]{\includegraphics[trim=77mm 1mm 77mm 5mm, clip, scale=0.45]{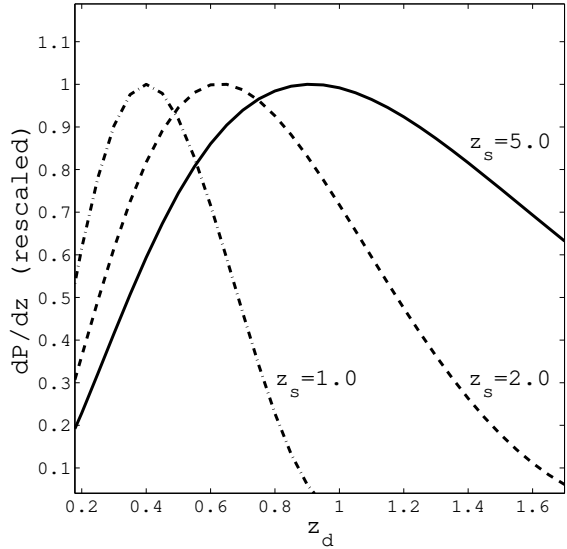}}
\caption{Lensing probability dP/dz rescaled to 1 for visual comparison of the peak. As sources move from $z_s=1.0$ to much higher redshift of $z_s=5.0$ the most dominant lens population redshift peak changes from $z_d\sim 0.4$ to $z_d\sim 1.0$.}
\label{f4}
\end{figure}
Plotting $dP/dz_d$ for a range of expected source redshifts is a good measure of the redshift distribution of lenses. Figure \ref{f4} shows that for sources beyond $z_s=1.0$ a significant population of lenses are located at redshifts of $z_d\sim 0.4$ to $z_d\sim 1.0$.

\subsection{Lensed Number Counts}
Gravitational lensing conserves the surface brightness of the lensed sources. Consequently any sources magnified by a factor of $\mu$ are also $\mu$ times brighter. 
In addition to making sources appear brighter, gravitational lensing also dilutes the source populations by magnifying the observed solid angles. Therefore it also dilutes the number counts by a factor of $\mu$. 

As discussed extensively in Jain and Lima (2009), these effects can
be combined for a large survey  to obtain the observed number counts as
\begin{equation}
\frac{dn}{dS}=\int \int \frac{1}{\mu'} \frac{dP}{d\mu'}(\mu',z) \: \frac{d\hat{n}}{d\hat{S}} (\hat{S}=\frac{S}{\mu'},z) \: dz \: d\mu'
\end{equation}
where we denoted the observed flux as S and the unlensed flux as $\hat{S}$,
such that $S=\mu \hat{S}$ and the unlensed differential source sky density
as $d\hat{n}/d\hat{S}$. 

\begin{figure}[h]
\centering
\makebox[0cm]{\includegraphics[trim=80mm 25mm 97mm 10mm, clip, scale=0.6]{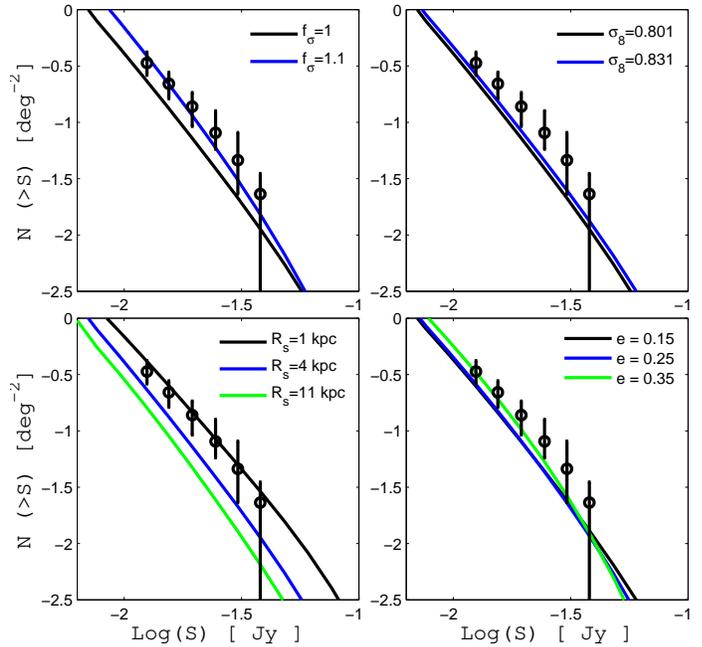}}
\caption{Effects of single parameter variations around a fiducial model ($f_{\sigma}=1$, $\sigma_8=0.801$,
$R_s=4 \,kpc$, and $e=0.15$) on predicted number counts of Durham unlensed model: $\sigma_{\nu}(M)$ (top left), $\sigma_8$ (top
right), source size (bottom left) and ellipticity (bottom right).}
\label{f5}
\end{figure}

The four panels in Figure \ref{f5} 
show the lensed count predictions for SPT as the most relevant parameters
are varied. 
The cosmological uncertainty is clearly subdominant to uncertainties in the
source properties and lens properties. In particular, the overall 
normalization of the mass profiles ($f_{\sigma}$) 
and the source size can change the expected number of strong lenses by 
large amounts.

Figure \ref{f6} and \ref{f7} illustrate that there are 
lensing models which
fit the SPT source counts quite well for both the Durham and UBC unlensed 
source count models. The lensing parameters are slightly different,
with the Durham model preferring slightly larger source sizes
and smaller ellipticity. 
The contribution of various redshift ranges to the lensed number counts 
can also be seen to be slightly different for the two models. Figure
\ref{f6} shows that the Durham
model finds that most of the sources lie at $z \ga 3$ over the entire
SPT flux range, while Figure \ref{f7} shows that the UBC model
has a significant fraction of lensed sources as low as $z\sim 2$. 

%The bottom plot shows the ratio of sources at each redshift 
%range to the total observed counts. In both cases, the lensed sources
%lie almost exclusively at $z \ga 2$.

\begin{figure}[h]
\centering
\makebox[0cm]{\includegraphics[trim=7mm 1mm 7mm 5mm, clip, scale=0.45]{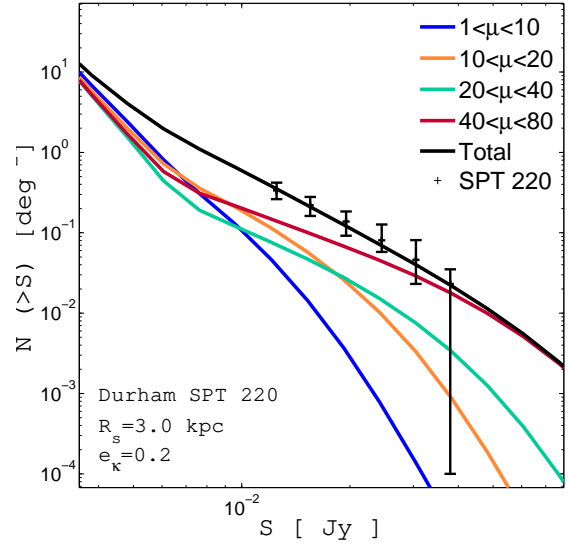}}
\caption{Magnification breakdown of the predicted SPT 220 GHz lensed counts for Durham model.
\\
\\
\\
\\
}
\label{f8}
\end{figure}

\begin{figure}[h]
\centering
\makebox[0cm]{\includegraphics[trim=7mm 1mm 7mm 5mm, clip, scale=0.45]{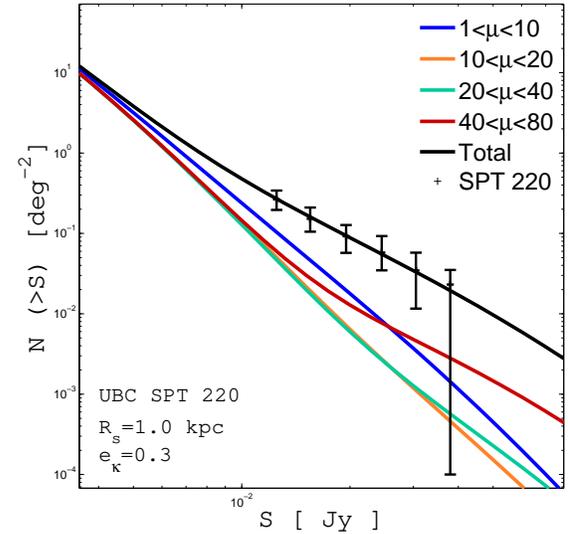}}
\caption{Same as Fig. \ref{f8} for UBC model.}
\label{f9}
\end{figure}

In Figure \ref{f8} and \ref{f9} the distribution in magnification can be seen to be a strong function of flux
for both models. At the low-flux end the cross-section
is dominated by relatively low magnifications ($\sim 10$), while 
higher fluxes are increasingly dominated by larger magnifications. This
is not surprising, given the steep unlensed luminosity function. 

The mean magnification at different fluxes is seen to have significant
differences between the Durham and UBC models, suggesting that this could
be a useful diagnostic for reconstructing the unlensed luminosity function.
In Figure \ref{f8} the Durham model is largely 
dominated by the highest magnifications in the region of the SPT counts,
while Figure \ref{f9} shows that the UBC model
shows a transition from low magnification to high magnification in the
flux range where SPT has reported constraints. This is a direct reflection
of the differences in the unlensed counts: the Durham model has an 
abrupt fall-off at the high flux end of the unlensed counts, well below
the flux range probed by SPT, while the
UBC model has more unlensed high redshift bright objects. 

\section{Discussion and Conclusions}

%\begin{figure*}
%\begin{center}
%\epsscale{1.4}
%\plottwo{FIG5-1.eps}{FIG5-2.eps}
%\caption{Best-fit lensed counts for SPT 220 GHz for Durham (left) and UBC (right) models. The Durham lensing consists of halos with $e_{\kappa}=0.2$ and $R_s=3 kpc$ while the UBC model's parameters are $e_{\kappa}=0.3$ and $R_s=1 kpc$. The colored curves show the contribution of each redshift bin to the total counts and the bottom panels plot the ratio of redshift bins to total counts. The unlensed counts only include sources with $z>0.2$ and the data points are SPT 220 dust counts with IRAS counterparts removed.}
%\label{DurhamREDSHIFT}
%\end{center}
%\end{figure*}

\begin{figure}[h]
\centering
\makebox[0cm]{\includegraphics[trim=7mm 1mm 7mm 5mm, clip, scale=0.45]{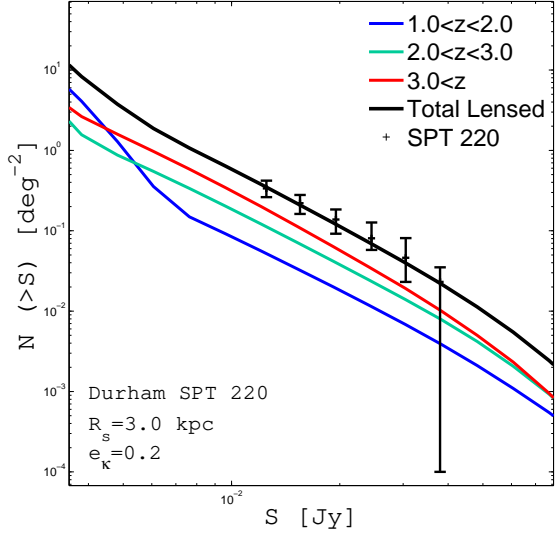}}
\caption{Best-fit lensed counts for SPT 220 GHz for Durham model. The lensing consists of halos with $e_{\kappa}=0.2$ and $R_s=3 kpc$. The colored curves show the contribution of each redshift bin to the total counts. The unlensed counts only include sources with $z>0.2$ and the data points are SPT 220 dust counts with IRAS counterparts removed.}
\label{f6}
\end{figure}

\begin{figure}[h]
\centering
\makebox[0cm]{\includegraphics[trim=7mm 1mm 7mm 5mm, clip, scale=0.45]{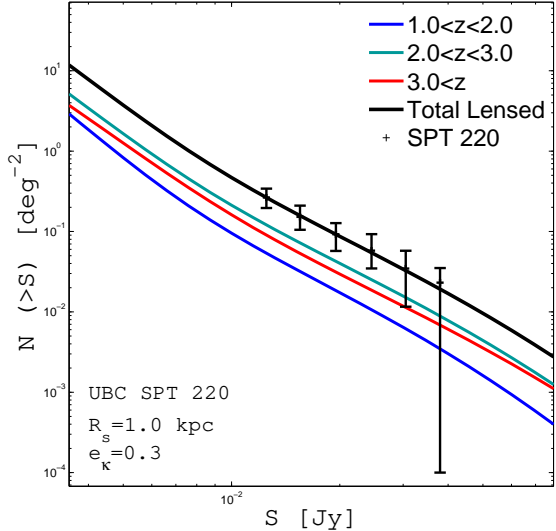}}
\caption{Same as Fig. \ref{f6} for UBC model with best fit parameters $e_{\kappa}=0.3$ and $R_s=1 kpc$. }
\label{f7}
\end{figure}

\begin{figure}[h]
\centering
\makebox[0cm]{\includegraphics[trim=7mm 1mm 7mm 5mm, clip, scale=0.45]{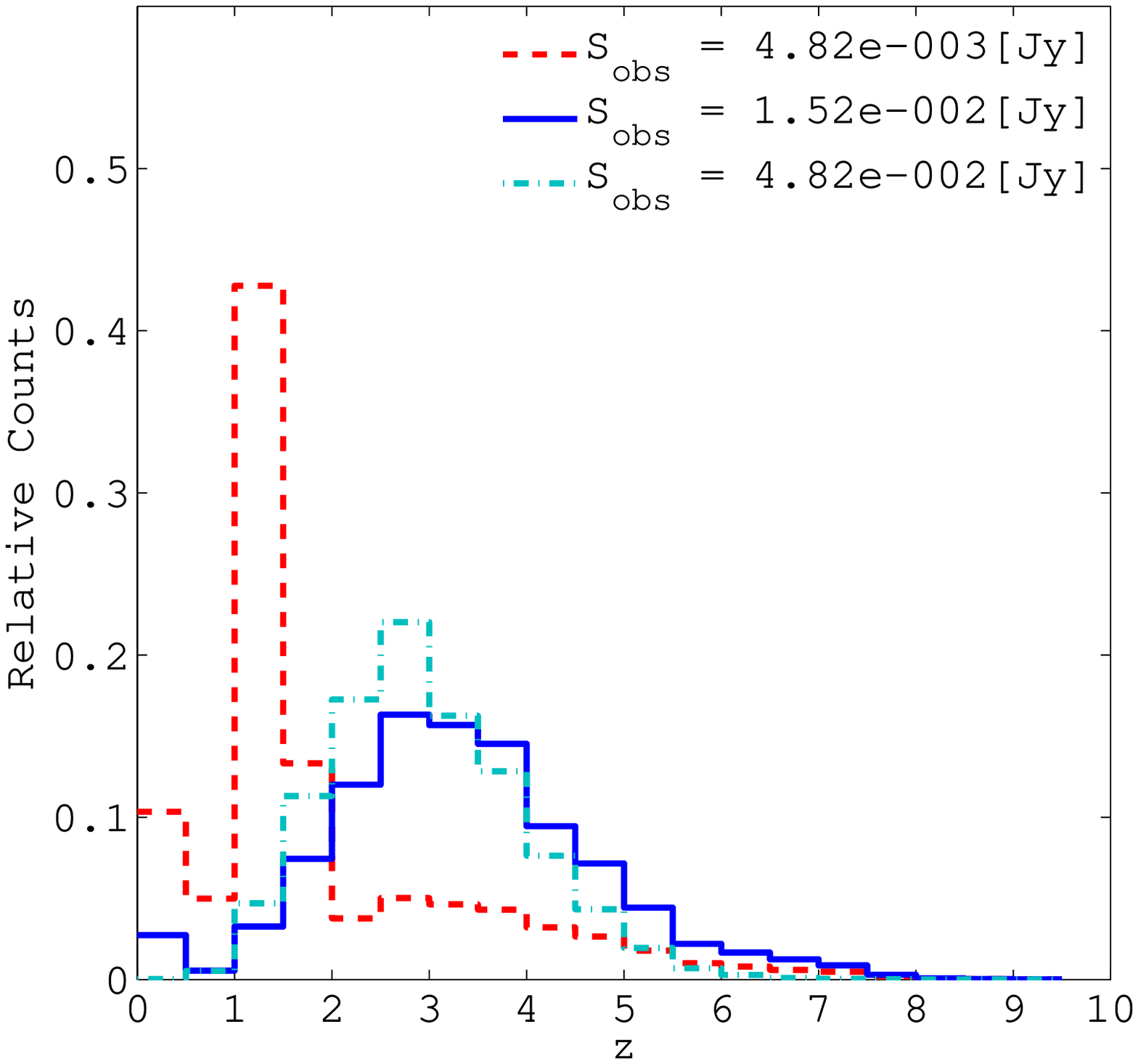}}
\caption{Normalized redshift distribution of the lensed sources at different observed flux levels. The model consists of Durham SPT 220 counts with $z>0.2$ lensed with halos with an ellipticity of $e_{\kappa}=0.2$ and sources with a radius of 3.0 kpc.}
\label{f10}
\end{figure}

The observed number counts of galaxies at mm wavelengths are
easily explained by gravitational lensing. However, 
the predictions for the number counts have several large sources
of uncertainty, several of which we have explored in some detail.

%%{Source Size}
The physical size of sources is an important factor in determining the 
amount of magnification. This has been an
important sources of uncertainty in previous calculations \citep{Paciga:09}. 
Mathematical point sources can be treated analytically, but for any 
realistically extended source the magnification can be strongly
affected. 
For large sources (compared to the lens Einstein radius) 
the maximum magnification is considerably reduced. As the relevant
scale is the Einstein radius (a property of the lens), this
effect will depend strongly on the mass of the lens. 
At intermediate magnifications (i.e. $10<\mu<40$) finite source effects 
can {\em increase} the lensing cross-section because larger sources have higher probability of having a part of them fall within the inner caustics and be 
highly magnified. 

For the
Durham model we observe that allowing sources with radius larger than 5 kpc 
leads to predicted counts falling below the observations. 
While it is not expected that the
massively star-forming region in these sources 
is significantly larger than this, this is another observational clue to
the physical processes in these galaxies. 
This has already been measured observationally for individual sources \citep{Momjian:10}. More extensive future follow-up observation of more SMG lensed sources could give better observational constraints on the range of SMG sizes. In addition it is possible to make a superposition of the curves of number counts for different assumed source sizes in order to achieve total counts for a diverse population of source sizes. This is mainly relevant to the distinction of AGN sources from star formation regions which extend over much larger spatial extents. For example assuming a $30\%$ AGN contribution to the total counts one can superpose the curve of counts for point sources (AGN) and extended sources (star formation) with the weight of 0.3 and 0.7. Nevertheless this work shows that the effects of other uncertainties in modeling the lens (e.g., halo velocity dispersion normalization) on the counts dominate the more subtle effects such as the distribution of source sizes.

Ellipticity is important in two ways: it increases the total area of high
magnification by increasing the area of the so-called ``diamond caustic,'' 
but it also leads to more sharply-defined structures in the
source plane, making finite source size effects more important. 
The trade-off between these effects was found to be an important effect,
as this can change the local slope of the lensed number counts.

A more realistic approach would be to assume a distribution of 
ellipticities and source sizes to constrain the parameter space. However,
with the strong parameter degeneracies we have identified it is clear
that more data will be required to separate these effects.

%%{redshift distribution}
There remains significant uncertainty in the expected redshift
distribution of the sources that comprise the mm-wave background from
star-forming galaxies. Gravitational lensing provides a magnified view
of these objects, providing an opportunity to probe the star formation
history. 
As a measure of the source redshift distribution we have plotted (Figure \ref{f10})
the redshift distribution of the lensed counts (normalized to 1 for 
visual purposes) for three different observed flux levels.
The redshift distribution of lensed sources at each observed flux level depends 
both on the redshift distribution of the unlensed model and the applied 
lensing probability function. 

Source redshift distribution is an important indicator of the nature of submm populations. The recent SPT observations by \citet{Vieira:10} show a large population of mm sources which were previously unseen in other catalogs. 
These sources are speculated to be high-redshift gravitationally lensed 
galaxies. Our lensed number count plots suggest that this can be a 
possible scenario. 
Figure \ref{f6} and \ref{f7} show that sources with redshifts larger than $z>3$ (red curve) constitute at least half of the observed 
non-IRAS-detected dusty sources of \citet{Vieira:10}.

%%{Discussion}

Throughout this work we have assumed SIS profiles as the mass
model for lensing halos. Strong lensing is very sensitive to the inner structure of the halos very close to the center \citep{Mead:10}. Rotation
curves of galaxies which are approximately flat to large distances
\citep{Rubin:70} suggest that such a model is not a bad approximation to 
the mass density profile of galaxy-sized halos. 
While this is not likely to be a bad approximation even up to cluster scales,
the normalization of the mass density profile may not be expected to
follow the self-similar scaling expected for the large-scale velocity
dispersion of the dark halo. Given the strong sensitivity to the
overall normalization of the density profile seen in Figure
\ref{f5}, strong lensing number counts will be sensitive to
the details of the radial profiles of the mass density and its
evolution with mass. 

A more realistic calculation would involve taking an NFW halo 
\citep{NFW:97}, correcting for baryonic condensation at the center,
 and populating it with smaller subhalos according to galaxy 
occupation numbers \citep{Oguri:06}. 
The positioning of the subhalos could be an important factor, as would
the density profiles of the substructure. A 
similar work was done by \citet{Meneghetti:03} where the effects of a 
cD galaxy on the cross-section of the main halo have been examined. 
This could perhaps be better resolved by carrying out ray-tracing 
simulations through N-body simulations which include realistic baryonic matter. 
Such a work has been carried out by \citet{Hilbert:07} where they study 
ray-tracing through the Millennium simulation; they include baryonic 
matter based on semi-analytic models of star formation in a related study 
\citep{Hilbert:08}. These studies have not had the required 
resolution at galaxy-scale levels to study strong lensing due to substructure.  
In general, there is very little empirical guidance at this time
for connecting Einstein radius to halo mass for halos smaller than
galaxy clusters; population studies of strong lenses may provide
some of the best constraints.

Since the SPT does not have the power to resolve multiple images of strongly lensed 
sources, the magnification of individual images does not affect the observations. Hence in
our calculations of magnification we have summed over the area of all the images of each source. 
If future observations succeed to resolve individual images then the multiplicity of the images 
(e.g. the ratio of quadruply lensed to doubly lensed images) can put stronger constraints on the
lens parameters, in particular lens ellipticity. The ratio of quadruply to doubly lensed sources can be obtained from the ratio of the area of the diamond caustic to the radial caustic \citep{Kormann:94} which strongly depends on the ellipticity of the lens. There has however been a strong disagreement between the observed ratio of the quad to double images and theoretical calculations \citep{Rusin:02}. One of the possible effects causing this discrepancy may be observational selection effects due to the higher magnification of the quads. For example we calculate that the ratios of $\mu>=10$ cross-section (a typical magnification of SPT sources based on our work) to the area of the diamond caustic for ellipticities of $e_{\kappa}=0.1$, $e_{\kappa}=0.2$ and $e_{\kappa}=0.3$ are $4.12\%$, $19.24\%$, and $49.28\%$ respectively. However perhaps the issue of quad to double ratios requires a more involved analysis which is beyond the scope of this work.

The methods used here can be generally applied to number count 
predictions for observations with other instruments such as 
Herschel (operating at 250, 350, and 500 $\mu$m), BLAST, etc. 
The large sky coverage of SPT makes it particularly efficient, probing
the rarest objects, where a large fraction are strongly lensed. On the other
hand, the deeper Herschel observations allow counts to lower fluxes, allowing
better characterization of the unlensed source distribution. 
A combined analysis of Herschel and SPT lensed counts will provide key 
insights into properties of both gravitational lenses and star-forming 
galaxies at high redshift. The results of \citet{Vieira:10} were based on
less than $10\%$ of the final projected SPT survey area, and the
Herschel observations are just now starting to emerge.
The upcoming Atacama Large Millimeter Array  will have the resolution
to make detailed images of these sources, providing more precise 
observational constraints on theoretical models presented here.

\acknowledgements{We thank Joaquin Vieira and Bhuvnesh Jain for useful
conversations, and Cedric Lacey and Gaelen Marsden for providing 
unpublished source count models at SPT wavelengths. This work was
supported by the Canadian Institute for Advanced Research, as well as
NSERC Discovery and the Canada Research Chairs program.}

%\bibHeading{References}
\bibliography{references}

\begin{thebibliography}{38}
\expandafter\ifx\csname natexlab\endcsname\relax\def\natexlab#1{#1}\fi

\bibitem[{{Bartelmann} \& {Schneider}(2001)}]{Bartelmann:01}
{Bartelmann}, M., \& {Schneider}, P. 2001, \physrep, 340, 291

\bibitem[{{Baugh} {et~al.}(2005){Baugh}, {Lacey}, {Frenk}, {Granato}, {Silva},
  {Bressan}, {Benson}, \& {Cole}}]{Baugh:05}
{Baugh}, C.~M., {Lacey}, C.~G., {Frenk}, C.~S., {Granato}, G.~L., {Silva}, L.,
  {Bressan}, A., {Benson}, A.~J., \& {Cole}, S. 2005, \mnras, 356, 1191

\bibitem[{{Blain}(1998)}]{Blain:98}
{Blain}, A.~W. 1998, \mnras, 297, 511

\bibitem[{{Blain} {et~al.}(2002){Blain}, {Smail}, {Ivison}, {Kneib}, \&
  {Frayer}}]{Blain:02}
{Blain}, A.~W., {Smail}, I., {Ivison}, R.~J., {Kneib}, J., \& {Frayer}, D.~T.
  2002, \physrep, 369, 111

\bibitem[{{Bryan} \& {Norman}(1998)}]{Bryan:98}
{Bryan}, G.~L., \& {Norman}, M.~L. 1998, \apj, 495, 80

\bibitem[{{Coppin} {et~al.}(2006){Coppin}, {Chapin}, {Mortier}, {Scott},
  {Borys}, {Dunlop}, {Halpern}, {Hughes}, {Pope}, {Scott}, {Serjeant}, {Wagg},
  {Alexander}, {Almaini}, {Aretxaga}, {Babbedge}, {Best}, {Blain}, {Chapman},
  {Clements}, {Crawford}, {Dunne}, {Eales}, {Edge}, {Farrah}, {Gazta{\~n}aga},
  {Gear}, {Granato}, {Greve}, {Fox}, {Ivison}, {Jarvis}, {Jenness}, {Lacey},
  {Lepage}, {Mann}, {Marsden}, {Martinez-Sansigre}, {Oliver}, {Page},
  {Peacock}, {Pearson}, {Percival}, {Priddey}, {Rawlings}, {Rowan-Robinson},
  {Savage}, {Seigar}, {Sekiguchi}, {Silva}, {Simpson}, {Smail}, {Stevens},
  {Takagi}, {Vaccari}, {van Kampen}, \& {Willott}}]{Coppin:06}
{Coppin}, K., {et~al.} 2006, \mnras, 372, 1621

\bibitem[{{Evans} \& {Bridle}(2009)}]{Evans:09}
{Evans}, A.~K.~D., \& {Bridle}, S. 2009, \apj, 695, 1446

\bibitem[{{Fedeli} \& {Berciano Alba}(2009)}]{Fedeli:09}
{Fedeli}, C., \& {Berciano Alba}, A. 2009, \aap, 508, 141

\bibitem[{{Frayer} {et~al.}(2010){Frayer}, {Harris}, {Baker}, {Ivison},
  {Smail}, {Negrello}, {Maddalena}, {Amblard}, {Auld}, {Baes}, {Birkinshaw},
  {Bonfield}, {Burgarella}, {Buttiglione}, {Cava}, {Clements}, {Cooray},
  {Dannerbauer}, {Dariush}, {De Zotti}, {Dunlop}, {Dunne}, {Dye}, {Eales},
  {Fritz}, {Gonzalez-Nuev}, {Herranz}, {Hopwood}, {Ibar}, {Jarvis}, {Lagache},
  {Leeuw}, {Lopez-Caniego}, {Maddox}, {Michalowski}, {Omont}, {Pascale},
  {Pohlen}, {Rigby}, {Rodighiero}, {Samui}, {Scott}, {Serjeant}, {Sibthorpe},
  {Smith}, {Swinbank}, {Temi}, {Thompson}, {Valtchanov}, {van der Werf}, \&
  {Verma}}]{Frayer:10}
{Frayer}, D.~T., {et~al.} 2010, ArXiv e-prints

\bibitem[{{Hall} {et~al.}(2009){Hall}, {Knox}, {Reichardt}, {Ade}, {Aird},
  {Benson}, {Bleem}, {Carlstrom}, {Chang}, {Cho}, {Crawford}, {Crites}, {de
  Haan}, {Dobbs}, {George}, {Halverson}, {Holder}, {Holzapfel}, {Hrubes},
  {Joy}, {Keisler}, {Lee}, {Leitch}, {Lueker}, {McMahon}, {Mehl}, {Meyer},
  {Mohr}, {Montroy}, {Padin}, {Plagge}, {Pryke}, {Ruhl}, {Schaffer}, {Shaw},
  {Shirokoff}, {Spieler}, {Staniszewski}, {Stark}, {Switzer}, {Vanderlinde},
  {Vieira}, {Williamson}, \& {Zahn}}]{Hall:09}
{Hall}, N.~R., {et~al.} 2009, ArXiv e-prints

\bibitem[{{Hilbert} {et~al.}(2007){Hilbert}, {White}, {Hartlap}, \&
  {Schneider}}]{Hilbert:07}
{Hilbert}, S., {White}, S.~D.~M., {Hartlap}, J., \& {Schneider}, P. 2007,
  \mnras, 382, 121

\bibitem[{{Hilbert} {et~al.}(2008){Hilbert}, {White}, {Hartlap}, \&
  {Schneider}}]{Hilbert:08}
---. 2008, \mnras, 386, 1845

\bibitem[{{Jain} \& {Lima}(2010)}]{Jain:10}
{Jain}, B., \& {Lima}, M. 2010, ArXiv e-prints

\bibitem[{{Keeton}(2001)}]{Keeton:01}
{Keeton}, C.~R. 2001, ArXiv Astrophysics e-prints

\bibitem[{{Koopmans} {et~al.}(2009){Koopmans}, {Bolton}, {Treu}, {Czoske},
  {Auger}, {Barnab{\`e}}, {Vegetti}, {Gavazzi}, {Moustakas}, \&
  {Burles}}]{Koopmans:09}
{Koopmans}, L.~V.~E., {et~al.} 2009, \apjl, 703, L51

\bibitem[{{Kormann} {et~al.}(1994){Kormann}, {Schneider}, \&
  {Bartelmann}}]{Kormann:94}
{Kormann}, R., {Schneider}, P., \& {Bartelmann}, M. 1994, \aap, 284, 285

\bibitem[{{Lacey} {et~al.}(2010){Lacey}, {Baugh}, {Frenk}, {Benson}, {Orsi},
  {Silva}, {Granato}, \& {Bressan}}]{Lacey:10}
{Lacey}, C.~G., {Baugh}, C.~M., {Frenk}, C.~S., {Benson}, A.~J., {Orsi}, A.,
  {Silva}, L., {Granato}, G.~L., \& {Bressan}, A. 2010, \mnras, 405, 2

\bibitem[{{Lima} {et~al.}(2009){Lima}, {Jain}, \& {Devlin}}]{Lima:09}
{Lima}, M., {Jain}, B., \& {Devlin}, M. 2009, ArXiv e-prints

\bibitem[{{Lima} {et~al.}(2010){Lima}, {Jain}, {Devlin}, \&
  {Aguirre}}]{Lima:10}
{Lima}, M., {Jain}, B., {Devlin}, M., \& {Aguirre}, J. 2010, ArXiv e-prints

\bibitem[{{Ludlow} {et~al.}(2010){Ludlow}, {Navarro}, {Springel},
  {Vogelsberger}, {Wang}, {White}, {Jenkins}, \& {Frenk}}]{Ludlow:10}
{Ludlow}, A.~D., {Navarro}, J.~F., {Springel}, V., {Vogelsberger}, M., {Wang},
  J., {White}, S.~D.~M., {Jenkins}, A., \& {Frenk}, C.~S. 2010, \mnras, 406,
  137

\bibitem[{{Marsden} {et~al.}(2010){Marsden}, {Chapin}, {Halpern}, {Patanchon},
  {Scott}, {Truch}, {Valiante}, {Viero}, \& {Wiebe}}]{Marsden:10}
{Marsden}, G., {et~al.} 2010, ArXiv e-prints

\bibitem[{{Mead} {et~al.}(2010){Mead}, {King}, {Sijacki}, {Leonard},
  {Puchwein}, \& {McCarthy}}]{Mead:10}
{Mead}, J.~M.~G., {King}, L.~J., {Sijacki}, D., {Leonard}, A., {Puchwein}, E.,
  \& {McCarthy}, I.~G. 2010, \mnras, 406, 434

\bibitem[{{Meneghetti} {et~al.}(2005){Meneghetti}, {Bartelmann}, {Dolag},
  {Perrotta}, {Baccigalupi}, {Moscardini}, \& {Tormen}}]{Meneghetti:05}
{Meneghetti}, M., {Bartelmann}, M., {Dolag}, K., {Perrotta}, F., {Baccigalupi},
  C., {Moscardini}, L., \& {Tormen}, G. 2005, NewAR, 49, 111

\bibitem[{{Meneghetti} {et~al.}(2003){Meneghetti}, {Bartelmann}, \&
  {Moscardini}}]{Meneghetti:03}
{Meneghetti}, M., {Bartelmann}, M., \& {Moscardini}, L. 2003, \mnras, 340, 105

\bibitem[{{Micha{\l}owski} {et~al.}(2010){Micha{\l}owski}, {Hjorth}, \&
  {Watson}}]{Michalowski:10}
{Micha{\l}owski}, M., {Hjorth}, J., \& {Watson}, D. 2010, \aap, 514, A67+

\bibitem[{{Momjian} {et~al.}(2010){Momjian}, {Wang}, {Knudsen}, {Carilli},
  {Cowie}, \& {Barger}}]{Momjian:10}
{Momjian}, E., {Wang}, W., {Knudsen}, K.~K., {Carilli}, C.~L., {Cowie}, L.~L.,
  \& {Barger}, A.~J. 2010, \aj, 139, 1622

\bibitem[{{Navarro} {et~al.}(1997){Navarro}, {Frenk}, \& {White}}]{NFW:97}
{Navarro}, J.~F., {Frenk}, C.~S., \& {White}, S.~D.~M. 1997, \apj, 490, 493

\bibitem[{{Negrello} {et~al.}(2007){Negrello}, {Perrotta},
  {Gonz{\'a}lez-Nuevo}, {Silva}, {de Zotti}, {Granato}, {Baccigalupi}, \&
  {Danese}}]{Negrello:07}
{Negrello}, M., {Perrotta}, F., {Gonz{\'a}lez-Nuevo}, J., {Silva}, L., {de
  Zotti}, G., {Granato}, G.~L., {Baccigalupi}, C., \& {Danese}, L. 2007,
  \mnras, 377, 1557

\bibitem[{{Oguri}(2006)}]{Oguri:06}
{Oguri}, M. 2006, \mnras, 367, 1241

\bibitem[{{Paciga} {et~al.}(2009){Paciga}, {Scott}, \& {Chapin}}]{Paciga:09}
{Paciga}, G., {Scott}, D., \& {Chapin}, E.~L. 2009, \mnras, 395, 1153

\bibitem[{{Perrotta} {et~al.}(2002){Perrotta}, {Baccigalupi}, {Bartelmann}, {De
  Zotti}, \& {Granato}}]{Perrotta:02}
{Perrotta}, F., {Baccigalupi}, C., {Bartelmann}, M., {De Zotti}, G., \&
  {Granato}, G.~L. 2002, \mnras, 329, 445

\bibitem[{{Press} \& {Schechter}(1974)}]{Press:74}
{Press}, W.~H., \& {Schechter}, P. 1974, \apj, 187, 425

\bibitem[{{Richard} {et~al.}(2010){Richard}, {Smith}, {Kneib}, {Ellis},
  {Sanderson}, {Pei}, {Targett}, {Sand}, {Swinbank}, {Dannerbauer}, {Mazzotta},
  {Limousin}, {Egami}, {Jullo}, {Hamilton-Morris}, \& {Moran}}]{Richard:10}
{Richard}, J., {et~al.} 2010, \mnras, 404, 325

\bibitem[{{Rubin} \& {Ford}(1970)}]{Rubin:70}
{Rubin}, V.~C., \& {Ford}, Jr., W.~K. 1970, \apj, 159, 379

\bibitem[{{Rusin} \& {Tegmark}(2001)}]{Rusin:02}
{Rusin}, D., \& {Tegmark}, M. 2001, \apj, 553, 709

\bibitem[{{Sheth} \& {Tormen}(1999)}]{Sheth:99}
{Sheth}, R.~K., \& {Tormen}, G. 1999, \mnras, 308, 119

\bibitem[{{Turner} {et~al.}(1984){Turner}, {Ostriker}, \& {Gott}}]{Turner:84}
{Turner}, E.~L., {Ostriker}, J.~P., \& {Gott}, III, J.~R. 1984, \apj, 284, 1

\bibitem[{{Vieira} {et~al.}(2010){Vieira}, {Crawford}, {Switzer}, {Ade},
  {Aird}, {Ashby}, {Benson}, {Bleem}, {Brodwin}, {Carlstrom}, {Chang}, {Cho},
  {Crites}, {de Haan}, {Dobbs}, {Everett}, {George}, {Gladders}, {Hall},
  {Halverson}, {High}, {Holder}, {Holzapfel}, {Hrubes}, {Joy}, {Keisler},
  {Knox}, {Lee}, {Leitch}, {Lueker}, {Marrone}, {McIntyre}, {McMahon}, {Mehl},
  {Meyer}, {Mohr}, {Montroy}, {Padin}, {Plagge}, {Pryke}, {Reichardt}, {Ruhl},
  {Schaffer}, {Shaw}, {Shirokoff}, {Spieler}, {Stalder}, {Staniszewski},
  {Stark}, {Vanderlinde}, {Walsh}, {Williamson}, {Yang}, {Zahn}, \&
  {Zenteno}}]{Vieira:10}
{Vieira}, J.~D., {et~al.} 2010, \apj, 719, 763

\end{thebibliography}
\bibliographystyle{apj}

\end{document}